\documentclass[prd,nofootinbib,preprint,superscriptaddress]{revtex4}

\usepackage{amsmath, amssymb, amsthm, graphicx, epsfig, fancyhdr,epsfig, slashed}

\usepackage{tikzsymbols}
\usepackage{natbib}
\usepackage{float}

\usepackage{tikz,xcolor,hyperref}

\usepackage{bm}
\usepackage{url}

\usepackage{natbib}

\usepackage{soul}

\newcommand{\be}{\begin{equation}}
\newcommand{\ee}{\end{equation}}
\newcommand{\bea}{\begin{eqnarray}}
\newcommand{\eea}{\end{eqnarray}}

\definecolor{lime}{HTML}{A6CE39}
\DeclareRobustCommand{\orcidicon}{
	\begin{tikzpicture}
	\draw[lime, fill=lime] (0,0) 
	circle [radius=0.2] 
	node[white] {{\fontfamily{qag}\selectfont \tiny ID}};
	\draw[white, fill=white] (-0.0625,0.095) 
	circle [radius=0.007];
	\end{tikzpicture}
	\hspace{-2mm}
}

\foreach \x in {A, ..., Z}{\expandafter\xdef\csname orcid\x\endcsname{\noexpand\href{https://orcid.org/\csname orcidauthor\x\endcsname}
			{\noexpand\orcidicon}}
}

\begin{document}

\title{Solutions for gravity coupled to Higgs--Yang-Mills dark sector with Jacobi elliptical functions}

\author{Marco Frasca\orcidA{}}
\email{marcofrasca@mclink.it}
\affiliation{Rome, Italy}

\author{Anish Ghoshal\orcidB{}}
\email{anish.ghoshal@fuw.edu.pl}
\affiliation{Institute of Theoretical Physics, Faculty of Physics, University of Warsaw, ul. Pasteura 5, 02-093 Warsaw, Poland}

\author{Massimiliano Rinaldi\orcidC{}}
\email{massimiliano.rinaldi@unitn.it}
\affiliation{Department of Physics, University of Trento, Via Sommarive 14, 38123 Povo (TN), Italy}
\affiliation{Trento Institute for Fundamental Physics and Applications-INFN, Via Sommarive 14, 38123 Povo (TN), Italy
}

\date{\today}

\begin{abstract}
\noindent We discuss the equations that arise from a Higgs--Yang-Mills dark sector coupled to gravity on a flat Friedmann-Lemaitre-Robinson-Walker metric. We choose the simplest $SU(2)$ representation, which we show to be compatible with the Cosmological Principle. We devise a multiple time scale approach to solve the equations of motion through a hierarchy of the couplings, utilizing exact solutions in terms of Jacobi elliptic functions. This novel method implements the dynamical system approach used in the literature and can shed new light on the possibility that this model can describe dark energy. 
\end{abstract}

\maketitle

\section{Introduction}

The interest in Einstein--Higgs--Yang-Mills (EHYM) theories in theoretical cosmology goes back to pioneering works by, among others, Henneaux \cite{Henneaux:1982vs}, Ochs et al. \cite{Ochs:1996yr}, and Galt'sov \cite{Galtsov:1991un,Galtsov:2008wkj}. 
More recently, the interplay of Higgs Yang-Mills fields with the cosmological dynamics has been studied in the context of inflation \cite{Adshead:2013nka,Adshead:2016omu,Adshead:2018emn,Maleknejad:2012fw,Maleknejad:2011jw}. Finally, EHYM equations were considered to build a model for dark energy (DE) in \cite{Rinaldi:2015iza,Alvarez:2019ues}. 

The common feature of these works is that EHYM equations can lead to accelerated expansion of the Universe without adding a cosmological constant or exotic matter content. Therefore, we believe that it is of great interest, both mathematically and physically, to find general solutions to the  EHYM equations in a cosmological context. 
With this aim in mind, we assume that the Higgs and Yang-Mills fields are part of a dark sector.

Generally speaking, the differential equations of motion resulting form EHYM gravity are of second order and non linear, hence there are no known exact solutions. In most of the works cited above, the equations are numerically solved. In the case of  \cite{Rinaldi:2015iza,Alvarez:2019ues} instead, the authors apply a dynamical system approach to find the fixed points and then study their stability, a practice that is well-known in the context of quintessence models of DE, see e.g. \cite{AmendolaBook}.
In the present work, we show an alternative, non-perturbative way to treat the Einstein-Higgs Yang-Mills (EHYM) equations coupled to a isotropic and homogeneous time-dependent metric.

More specifically, we exploit the exact solutions to the Higgs--Yang-Mills sector expressed in terms of Jacobi elliptical functions on the same footsteps of the quantum solution presented in Ref.\cite{Frasca:2015yva}, which were also also used in the case of quintessence Higgs dark energy \cite{Frasca:2022vvp}. 

We emphasize that our study is a first step and is purely classical. No attempt is made to evaluate quantum corrections since we do not commit to any theory of quantum gravity.  Black holes and compact objects have been thoroughly discussed in this context, see e.g. \cite{Hartmann:2001ic,Volkov:1998cc} and references therein. 
%

\textit{The paper is organized as follows:} in Sec.\ \ref{YMeqs} we review the equations of the Yang-Mills Higgs field theory minimally coupled to gravity. In Sec.\ \ref{nogauge} we consider the simplest case, that is when the gauge field vanishes, and solve the scalar field equations in terms of Jacobi functions to show the emergence of a mass gap. In Sec.\ \ref{yesgauge} we switch on the gauge fields and study the solutions to the field equations by extending the method explained in the previous section. In Sec.\ \ref{denergy} we show how the Higgs field, which is the only relevant one due to the hierarchy of the couplings, can act as dark energy, in line with the findings of \cite{Rinaldi:2015iza,Alvarez:2019ues}.
We then conclude in Sec.\ \ref{conc} with some remarks. Our work entails quite few mathematical steps that have been put in the appendices.

\section{Model}\label{YMeqs}

\noindent We consider the model proposed in \cite{Rinaldi:2015iza}. Here, it was suggested that dark energy might be sourced by the classical dynamics of the Yang-Mills-Higgs equations minimally coupled to gravity.  It was then pointed out in \cite{Alvarez:2019ues} that the choice of representation of the gauge group is critical and it was shown that the correct one is $SU(2)$. Thus, in the present work we consider the Lagrangian 
\be
\mathcal{L}=\sqrt{-\det g}\left(\frac{M^2_P}{2}R+\frac{1}{4}F_{\mu\nu}^aF^{a\mu\nu}-(D_\mu\Phi^a)^{\dagger}(D^\mu\Phi^a)+V(\Phi^2)\right)+\mathcal{L}_{\rm mat},
\ee
where $\Phi$ is the complex Higgs doublet
\bea
\Phi  = \begin{pmatrix} \phi_1 + i \psi_1 \\ \phi_2 + i \psi_2 \end{pmatrix}\,,
\eea 
\bea
F_{\mu\nu}^a =\partial_\mu A_\nu^a-\partial_\nu A_\mu^a+g\epsilon ^a_{\,\,bc}A^b_\mu A^c_\nu\,,
\eea
is the field strength expressed in terms of $A_\mu^a$ ($a=1,2,3$), which are the vector potential components.  The gauge coupling is $g$ and $\epsilon ^a_{\,\,bc}$ is the three-dimensional Levi-Civita tensor. 
\bea 
D_\mu=\partial_\mu-\frac12 g \tau^a A^a_\mu
\eea
is the covariant derivative for the gauge group, where $\tau^a$ are the $SU(2)$ generators. We choose the potential $V$ to be the standard Mexican-hat type
\bea
\color{red}V(\Phi^2)=\frac{\lambda}{4}(\left|\Phi\right|^2-\phi_0^2)^2\,,
\eea
where $\lambda$ is the coupling and $\phi_0$ the vacuum expectation value (v.e.v.) of the Higgs field. The term $\mathcal{L}_{\rm mat}$ denotes the eventual contribution of standard matter. Finally, $M_P$ is the Planck mass, $R$ the Ricci scalar and $\det g$ the determinant of the metric. 

In this work, we consider the Friedmann-Lemaitre-Robertson-Walker (FLRW) spatially flat metric
\be
ds^2=-dt^2+a^2(t)(dr^2+r^2d\Omega^2)\,,
\ee
since we have in mind possible applications to either inflationary cosmology or to models of dark energy.
By inspecting the Einstein equations adapted to this metric, it turns out that the only way to avoid anisotropic contributions to the stress tensor, which would be at odds with the diagonal structure of the metric, is to adopt the so-called cosmic triad, where the components of the gauge fields  reduce to one single dynamical degree of freedom $f(t)$, namely \cite{Alvarez:2019ues}
\be
\label{eq:Apot}
A^a_0=0\,, \quad A_i^a=f(t)\delta^a_i\,,
\ee
where $i$ denotes spatial indices. Essentially, this means that  the spatial components of the gauge field are the same in all directions (to preserve large-scale isotropy) and depend on time only (to preserve large-scale homogeneity). For consistency, the Higgs doublet also reduces to one single real component, that we call henceforth $\phi (t)$. Thus, the  potential now reads
\be
V(\phi)=\frac{\lambda}{4}(\phi^2-\phi_0^2)^2.
\ee
It is then straightforward to write down the equations of motion for the fields \cite{Alvarez:2019ues}. One has the standard Friedmann equations (for our aims, we can omit to consider matter and radiation contributions)
\bea\label{Hsq}
H^{2}&=&{1\over 3M^{2}_P}\left[{\dot\phi^{2}\over 2}+ {3\dot f^{2}\over 2a^{2}}+{2g^{2}f^{4}\over a^{4}}+ {3g^{2}f^{2}\phi^{2}\over 8a^2}+ V(\phi) \right]\,,\\\label{Hdot}
\dot H&=&-{1\over 2M^{2}_P}\left[\dot \phi^{2}+ {2\dot f^{2}\over a^{2}}+{2g^{2}f^{4}\over 4 a^{4}}+{g^{2}f^{2}\phi^{2}\over 4a^{2}} \right]\,, \nonumber 
\eea
where  $H=\dot a/a$. 
To these, one must add the two Klein-Gordon equations for $f$ and $\phi$, respectively
\bea\label{eqf}
\ddot f+H\dot f+{2g^{2}f^{3}\over a^{2}}+{g^{2}f\phi^{2}\over 4}=0\,,
\eea
and 
\bea\label{eqkg}
\ddot \phi+3H\dot\phi+{3g^{2}f^{2}\phi\over 4a^{2}}+\lambda\phi(\phi^{2}-\phi_{0}^{2})=0\,.
\eea
 In \cite{Rinaldi:2015iza} it was suggested that these equations might give an alternative explanation to the dark energy problem since the potential does not vanish, although it can be arbitrarily small. This is evident from eq.\  \eqref{eqkg}: here, the third term  prevents $\phi_0$ to be an exact solution of the equations of motion. It acts as a friction term that does not allow the Higgs field to reach the bottom of the potential in a finite time. This term vanishes only when either $g=0$ or $f/a\rightarrow 0$. As a result, in the first Friedmann equation the potential $V$ never vanishes and behaves as an effective cosmological constant, that might slowly vary in time according to the global dynamical evolutions of the equations of motion. The value of this effective cosmological constant at any time is set by the initial conditions of the dynamical system, whcih, however, must be very finely tuned, as shown in \cite{Rinaldi:2015iza}. 

In the present work, we consider again the equations above but instead of solving them in terms of a dynamical system as in \cite{Rinaldi:2015iza,Alvarez:2019ues}, we use the method described by one of us in \cite{Frasca:2015yva}, which alleviates the issue of initial conditions fine tuning.

\section{Case without gauge field}\label{nogauge}

\noindent To begin with, we consider the simplified case in which the gauge field $f$ is set to zero.This makes sense in view of the fact that we are not considering constant solutions and we expect a weak time evolution of the dark energy. Thus, the  equations above simplify to
\bea\label{eq:nogauge}
\ddot \phi+3H\dot\phi+\lambda\phi(\phi^{2}-\phi_{0}^{2})=0\,,
\eea
and 
\bea\label{Hsq1}
H^{2}&=&{1\over 3M^{2}_P}\left[{\dot\phi^{2}\over 2}+V(\phi)\right]\,,\\\label{Hdot1}
\dot H&=&-\frac{\dot \phi^{2}}{2M^{2}_P}\,. \nonumber
\eea
Now, we assume that the dynamics is ruled by two different time scales: the scalar field time-scale $t_\phi=\sqrt{\lambda}\phi_0t$  and the Hubble time-scale $\tau=3Ht$. 
We also define the new field 
\be
\chi(t)=e^{\frac{3}{2}Ht}\phi(t)\,.
\ee
As a result, the Klein-Gordon equation \eqref{eq:nogauge} can be written, approximately, as
\be
\ddot \chi-\frac{9}{4}H^2\chi+\lambda\chi(e^{-\tau}\chi^2-\phi_{0}^{2})=0\,,
\ee
where we assumed that the Hubble parameter changes very slowly compared to $\chi$. Now, we can apply a  multiple scale technique by assuming that $\chi=\chi(t_\phi,\tau)$. This requires a redefinition of the derivative as
\be
\frac{d}{dt}=\sqrt{\lambda}\phi_0\frac{\partial}{\partial t_\phi}+3H\frac{\partial}{\partial\tau}\,.
\ee
We make the perturbative expansion 
\be
\chi = \chi_0+\frac{1}{\sqrt{\lambda}}\chi_1+O(\lambda^{-1})\,,
\ee
to obtain the set of equations
\bea
&&\phi_0^2\frac{\partial^2}{\partial t_\phi^2}\chi_0+\chi_0(e^{-\tau}\chi_0^2-\phi_{0}^{2})=0\,, \nonumber \\
&&\phi_0^2\frac{\partial^2}{\partial t_\phi^2}\chi_1+(-\phi_0^2+3e^{-\tau}\chi_0^2)\chi_1
+6H\phi_0\frac{\partial^2}{\partial t_\phi\partial\tau}\chi_0=0\,, \nonumber \\
&&\vdots.
\eea
This is also known as the averaging method \cite{KevCole}. We expect that such a series should be at least asymptotic. With such an assumption, we are free to consider as range of validity $\color{red}\lambda \gtrsim 1$. The numerical example below confirms such expectation.

The leading order solution is given by
\be
\label{eq:chi01}
\chi_0=e^{\frac{\tau}{2}}\sqrt{\frac{2}{3}}\phi_0\,\text{dn}\left(\frac{1}{\sqrt{3}}t_\phi+\theta,-1\right),
\ee
where dn is a Jacobi elliptic function (see section 22 of Ref. \cite{NIST:DLMF} for further details)\footnote{We use the notation $\text{dn}(x,-1)$ because we work with the square of the modulus (the second argument of the function). Another notation is also possible using the modulus and, in such a case, one should write $\text{dn}(x,i)$ having identical numerical values. In Ref.~\cite{NIST:DLMF}, it is used the latter convention.} 
and $\theta$ an arbitrary integration constant. This function  never vanishes but oscillates with an amplitude  proportional to $\phi_0$. From a physical standpoint, this represents an oscillating background for the field that is never vanishing and so, it mimics a kind of Higgs dynamics. 
%
The frequency of the oscillations are proportional to the effective mass, which reads ${\phi_0}/{\sqrt{3}}$.

The next-to-leading order term is obtained by observing that the Green function $G(t_\phi,\tau)$, which solves the equation
\be
\phi_0^2\partial^2_{t_\phi}G(t_\phi,\tau)+(-\phi_0^2+3e^{-\tau}\chi_0^2(t_\phi,\tau))G(t_\phi,\tau)=\phi_0\delta(t_\phi)\,,
\ee
is given by
\be\label{eq:green}
G(t_\phi,\tau)=\frac{\sqrt{3}}{\sqrt{\lambda}\phi_0}\theta(t_\phi)\,\text{cn}\left(\frac{1}{\sqrt{3}}t_\phi,-1\right)\text{sn}\left(\frac{1}{\sqrt{3}}t_\phi,-1\right),
\ee
where cn and sn are standard Jacobi elliptic functions \cite{NIST:DLMF}. 

\begin{center}
\begin{figure}[H]
\includegraphics[height=8cm,width=8cm]{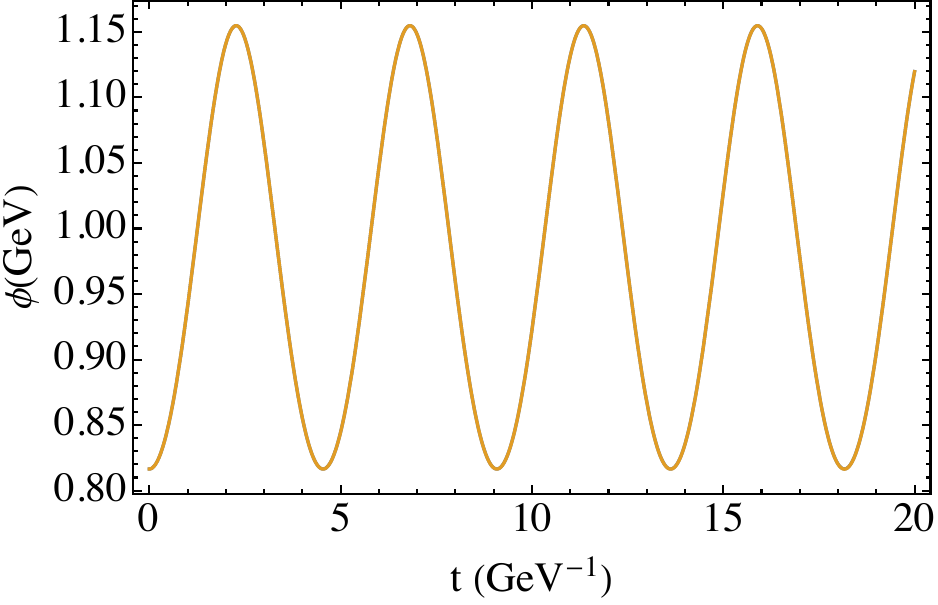}%
\includegraphics[height=8cm,width=8cm]{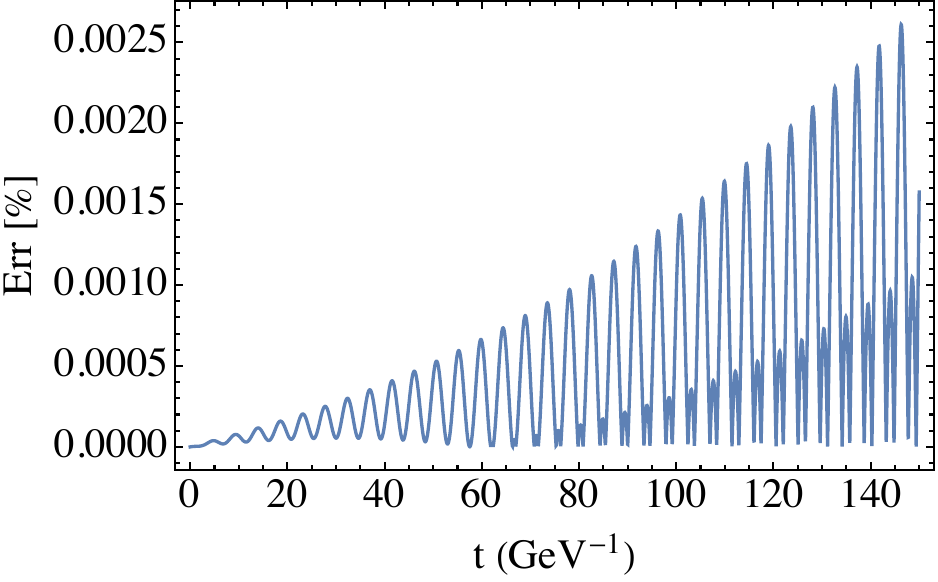}%
\caption{\it Plot on the \textbf{left panel} shows the behavior of the analytical solution given in eq.(\ref{eq:chi02}). We are assuming $\theta=0$ and $\dot\phi(0)=0$. The \textbf{right panel} shows the relative difference between the numerical solution of eq.(\ref{eq:nogauge}) and the analytical solution in eq.(\ref{eq:chi02}) in terms of percentage error at increasing time, given by Err $ = |1-\phi_{\rm numerical}/\phi_{\rm analytical}|$.\label{fig-1} 
}
\end{figure}
\end{center}

\begin{center}
\begin{figure}[H]
\includegraphics{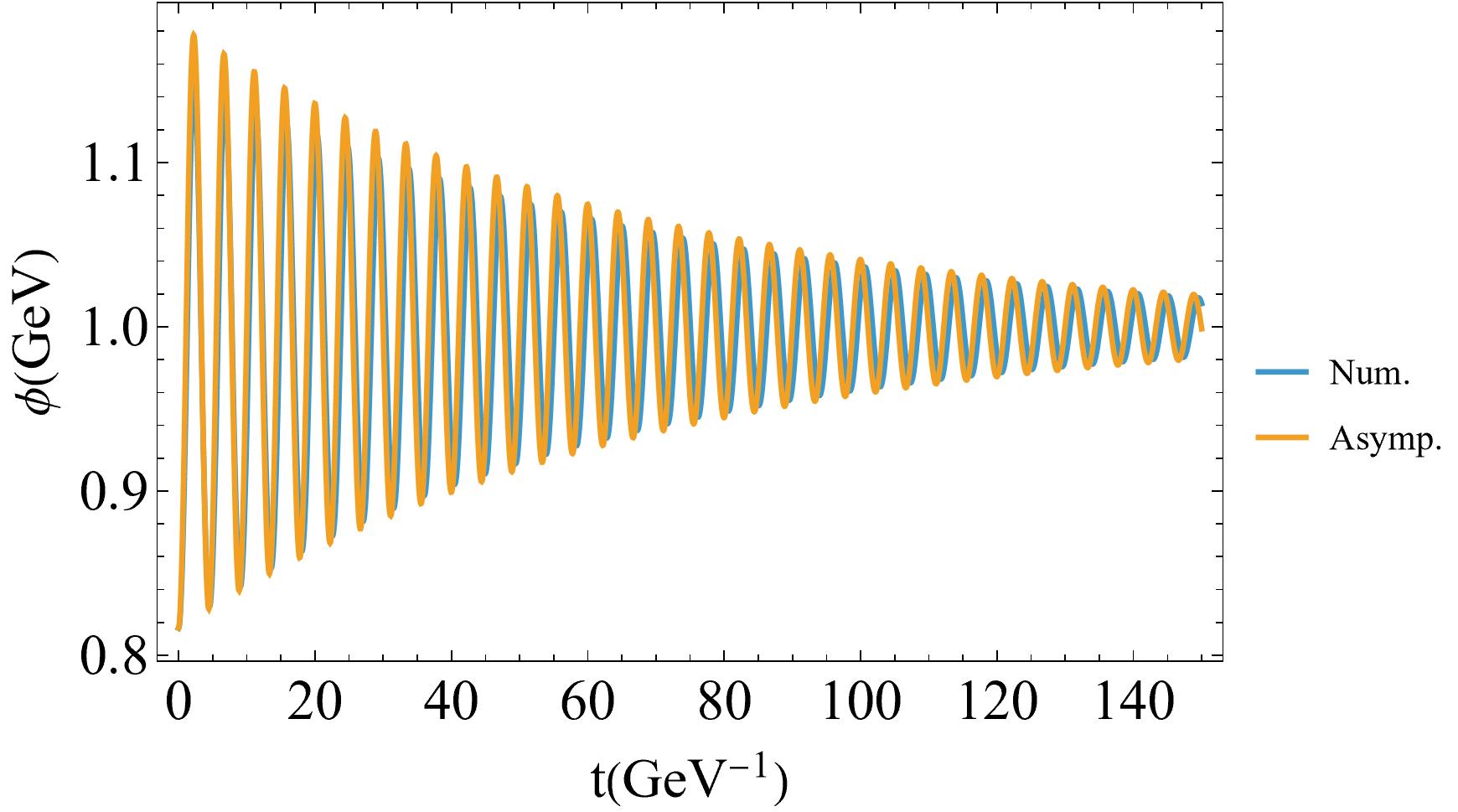}%
\caption{\it The behavior of the analytical solution given in eq.(\ref{eq:phia}) is compared to the numerical one with $H=0.01\ \rm{GeV}$, $\phi_0=1\ \rm{GeV}$ and $\lambda=1$. These values are chosen to make as clear as possible the comparison with the asymptotic solution.\label{fig-01} 
}
\end{figure}
\end{center}

We evaluate the first-order correction to the solution of the background in Appendix B, where we show that a secular term appears, producing a correction to  $\phi_0$ of the form
\be
\phi_{0R}(\tau)=\exp\left[{-K_v \tau/{\sqrt{6\lambda}}}\right]\phi_0.
\ee
where $K_v \approx 0.37$ is a constant. 
This yields a decay of the constant $\phi_0$ of the scalar field in eq.\ (\ref{eq:chi01}), at least for its amplitude.
The next-to-leading order solution is then
\be
\label{eq:chi02}
\phi(t)\approx \sqrt{\frac{2}{3}}\epsilon(\tau)\phi_0\,\text{dn}\left(\sqrt{\frac{\lambda}{3}}\phi_0t+\theta,-1\right),
\ee
where we have set $\epsilon(\tau)=\exp\left[{-K_v \tau/{\sqrt{6\lambda}}}\right]$, which describes the decay due to the time scale $\tau=3Ht$, which is ultimately related to the Hubble parameter.
This result essentially represents nonlinear oscillations damped by the expansion of the universe. To understand how good this approximation is, we have numerically solved eq.(\ref{eq:nogauge}) fixing the parameters to $H=10^{-5}\ \text{GeV}$, $\lambda=1$ and $\phi_0=1\ \text{GeV}$, just to be sure to have $\phi_0\gg H_0$. In Fig.~\ref{fig-1} we compare the numerical solution to the leading-order expression in eq.(\ref{eq:chi02}).
Condidering the asymptotic limit, we need to correct this solution with higher order terms to catch the right asymptotic value that is $\phi_0$, as it should be expected. A short proof of this statement is obtained by inserting in eq.(\ref{eq:nogauge}) the function
\be
\label{eq:asym}
\phi(t)=\phi_0+e^{-\frac{3}{2}Ht}\hat\phi(t),
\ee
which yields the equation
\be
\ddot{\hat\phi}+\left(2\lambda\phi_0^2-\frac{9}{4}H^2\right)\hat\phi+3e^{-\frac{3}{2}Ht}\lambda\phi_0\hat\phi^2+
e^{-3Ht}\lambda\hat\phi^3=0.
\ee
Then, the asymptotic limit $t\rightarrow\infty$ proves the result as  $\hat\phi$ is bounded. 
Thus, the asymptotic limit for $Ht\gtrapprox 1$ takes the form
\be
\label{eq:phia}
\phi(t)=\phi_0+e^{-\frac{3}{2}Ht}\left(\sqrt{\frac{2}{3}}-1\right)\phi_0\cos\left(\sqrt{2\lambda\phi_0^2-\frac{9}{4}H^2}t+\theta'\right),
\ee
where the phase $\theta'$ reconnects to the factor $\rm{dn}(\theta,-1)$ with eq.(\ref{eq:chi02}) at $t=0$. In Fig.~\ref{fig-01}, we compare the numerical and the analytical solutions and the result is quite satisfactory, being just the leading-order approximation.
As expected, the points $\pm\phi_0$ are asymptotically stable due to the Hubble constant.

We have found solutions describing the behavior of the scalar field for $Ht\ll 1$ and $Ht\gtrapprox 1$ but $\phi_0\gg H$. Thus, we can conclude that the solution in eq.(\ref{eq:chi02}) is well-suited when the scale factor $a$ is practically 1 that is the approximation $Ht\approx 0$. If we aim to study a situation where the scale of the Hubble constant starts to become important, then it is better to use eq.(\ref{eq:phia}). We are assuming that the scale of variation of the Hubble constant $H$ is the slowest possible in our study. From the Friedmann equations we have
\be
\label{eq:HL}
H^2\rightarrow H_L^2=\frac{\lambda\phi_0^4}{12M_P^2}, \qquad {\dot H}\rightarrow 0,
\ee
due to the damping factor. For the dark energy, we evaluate the equation of state as
\be
\label{eq:DEeos}
w=\frac{{\dot\phi^{2}\over 2}-V(\phi)}{{\dot\phi^{2}\over 2}+V(\phi)}.
\ee
Similarly to the Friedmann equations (\ref{Hsq1}) that give (\ref{eq:HL}), we have an asymptotic limit for eq.(\ref{eq:DEeos}) also yielding
\be
w\rightarrow -1,
\ee
for times large enough. Anyway, we need to evaluate the time dependence. We get 
\bea
\label{eq:w}
&&w=-1+\frac{8}{9}\epsilon^2(\tau)
\left[
\text{cn}\left(\frac{\phi_0\sqrt{\lambda}}{\sqrt{3}}t+\theta,-1\right)^2 \text{sn}\left(\frac{\phi_0\sqrt{\lambda}}{\sqrt{3}}t+\theta,-1\right)^2
\right. \\
&&-3\sqrt{2}\frac{K_vH}{\lambda\phi_0}\text{cn}\left(\frac{\phi_0\sqrt{\lambda}}{\sqrt{3}}t+\theta,-1\right)
\text{dn}\left(\frac{\phi_0\sqrt{\lambda}}{\sqrt{3}}t+\theta,-1\right) \text{sn}\left(\frac{\phi_0\sqrt{\lambda}}{\sqrt{3}}t+\theta,-1\right)+ \nonumber \\
&&\left.\frac{9}{2}\frac{K_v^2H^2}{\lambda^2\phi_0^2}
\text{dn}^2\left(\frac{\phi_0\sqrt{\lambda}}{\sqrt{3}}t+\theta,-1\right)\right]+O(\epsilon^4(\tau)). \nonumber
\eea
The dark energy equation of state parameter is time-varying but, due to the exponential damping arising from the expansion of the universe, it reaches asymptotically a constant value -1. This adds to the Friedmann equation for $\dot H$ yielding an accelerating term. This can be seen by considering the second Friedmann equation in the form
\be
\frac{\ddot a}{a}=-\frac{1}{3M_P^2}\left(\rho+3p\right)=-\frac{2}{3M_P^2}\left({\dot\phi}^2-V(\phi)\right),
\ee
that, in our case, is larger than 0 yielding, in the asymptotic limit,
\be
\frac{\ddot a}{a}=\frac{\lambda\phi_0^4}{6M_P^2}=2H_L^2.
\ee

\medskip

\section{Introducing Gauge fields}\label{yesgauge}

\noindent When a gauge field is present, we need to consider the set of equations (\ref{eqf})
\bea\label{eq:ff12}
&&\ddot f+H\dot f+{2g^{2}f^{3}\over a^{2}}+{g^{2}f\phi^{2}\over 4}=0\,, \nonumber \\
&&\ddot \phi+3H\dot\phi+{3g^{2}f^{2}\phi\over 4a^{2}}+\lambda\phi(\phi^{2}-\phi_{0}^{2})=0\,,
\eea
together with the Friedmann equations (\ref{Hsq}).

In the preceding section, we observed that the Hubble constant and the Higgs field run on different time scales, the latter much faster than the former. Now, due to the fact that the strong coupling is very large, towards the confining regime, the scale of the Yang-Mills sector 
should represent the fastest time scale. Similarly as before, we consider $H$ slowly varying and we redefine $f\rightarrow e^{-\frac{3}{2}Ht}f=\tilde f$ so that
\be
\ddot {\tilde f}-\frac{9}{4}H^2 {\tilde f}+e^{-3Ht}{2g^{2}{\tilde f}^{3}\over a^{2}}+{g^{2}{\tilde f}\phi^{2}\over 4}=0\,.
\ee
If we rescale the time as $t\rightarrow gt$, where $g$ is strong coupling constant, we find
\be
\ddot {\tilde f}-\frac{9}{4g^2}H^2 {\tilde f}+e^{-\frac{3}{g}Ht}{2{\tilde f}^{3}\over a^{2}}+{{\tilde f}\phi^{2}(t/g)\over 4}=0\,.
\ee
We just note that the evolution of the scale factor $a$ is on the same time-scale as H, the slowest one. In the formal limit $g\rightarrow\infty$, the exponential is practically inessential and $\phi$ can be very slowly varying, leaving us with the following leading order equation to solve
\be
\ddot {\tilde f}_0+e^{-\frac{3}{g}Ht}{2{\tilde f}_0^{3}\over a^{2}}+{{\tilde f}_0\phi^{2}(t/g)\over 4}=0\,.
\ee

The solution for $\phi$ is then given by eq.(\ref{eq:chi02}). Introducing the timescales $\tau=Ht$, $t_1=\phi_0t$, and removing the scaling in $g$, we obtain
\be
\ddot {\tilde f}_0(t)+e^{-3\tau}{2g^2{\tilde f}_0^{3}(t)\over a^{2}(\tau)}+{g^2{\tilde f}_0(t)\phi^{2}(t_1,\tau)\over 4}=0\,.
\ee
This equation admits the following exact solution
\bea
\label{eq:f00}
{\tilde f}_0&=&\sqrt{\frac{2\mu^4}{{g^2\phi^2(t_1,\tau)\over 4}+\sqrt{{g^4\phi^4(t_1,\tau)\over 16}+\left(e^{-3\tau}{4g^2\over a^{2}}\right)\mu^4}}}\times \\
&&\operatorname{sn}\left(m_0(t_1,\tau)t+\theta,
\sqrt{\frac{-{g^2\phi^2(t_1,\tau)\over 4}+\sqrt{{g^4\phi^4(t_1,\tau)\over 16}+\left(e^{-3\tau}{4g^2\over a^{2}}\right)\mu^4}}{-{g^2\phi^2(t_1,\tau)\over 4}-\sqrt{{g^4\phi^4(t_1,\tau)\over 16}+\left(e^{-3t_2}{4g^2\over a^{2}}\right)\mu^4}}}
\right), \nonumber
\eea
where $\mu$ and $\theta$ are integration constants, and 
\be
m_0^2(t_1,\tau)={g^2\phi^2(t_1,\tau)\over 4}+\frac{e^{-3\tau}{2g^2\over a^{2}}\mu^4}{{g^2\phi^2(t_1,\tau)\over 4}+\sqrt{{g^4\phi^4(t_1,\tau)\over 16}+\left(e^{-3\tau}{4g^2\over a^{2}}\right)\mu^4}}.
\ee
is the classical mass gap of the gauge field.
The last step is to reinstate the exponential, yielding 
\be
\label{eq:f0e}
f_0=e^{-(3/2)\tau}\tilde f_0.
\ee
We see that, on the $\tau$ time-scale, the scalar field becomes more rapidly irrelevant and we are left with the well-known solution for the Yang-Mills field with a mass gap, as shown in \cite{Frasca:2015yva}, multiplied by a decaying exponential on the $\tau$ timescale, namely
\be
\label{eq:f0}
f_0\approx e^{-\frac{3}{4}\tau}\mu\left(\frac{a}{g}\right)^\frac{1}{2}\operatorname{sn}(e^{-\frac{3}{2}\tau}(g/a)^\frac{1}{2}\mu t+\theta,i).
\ee
The phase $\theta$ can assume arbitrary values. An average on it yields $\langle f_0\rangle=0$, differently from the case of the scalar field. Besides, at large times, the gauge field decouples from gravity due to the growth of the scale factor. We emphasize that the decoupling is only with respect to gravity due to the hierarchy between the gauge and quartic couplings $g\gg\lambda$. The quantum theory of the gauge sector may display confinement but, at this classical level, we may only have a mass gap. Thus, the consistency with the phase scenario of a SU(N)-Higgs theory, along also with gravity, is not ruined.

\begin{center}
\begin{figure}[H]
\includegraphics[width=\textwidth]{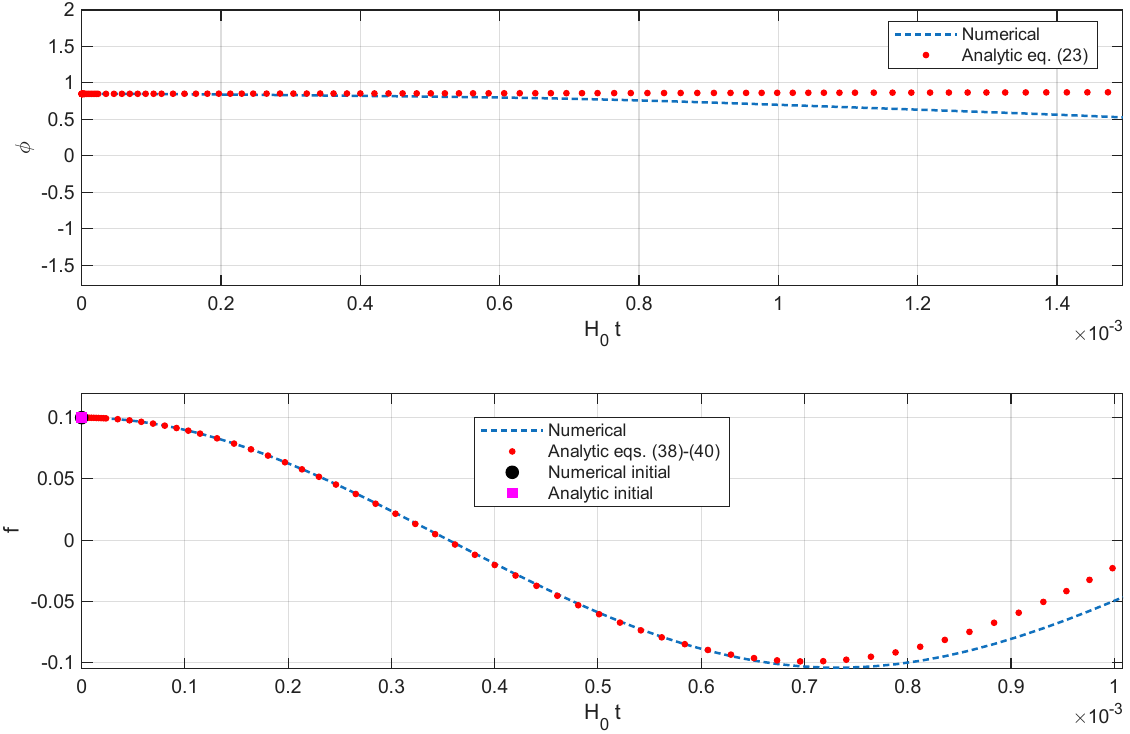}%
\caption{\it Numerical and analytical results compared for the gauge solution. Here $\color{red}H_0=10^{-5}\, \text{GeV}$. We evidence the excellent coincidence of the initial values for the numerical and the analytical solutions.\label{fig-g} 
}
\end{figure}
\end{center}

Aiming to have a straightforward confirmation of this scenario, we numerically solved eq.(\ref{eq:ff12}) and compared the results with eq.(\ref{eq:f0e}) and eq.(\ref{eq:chi02}) for the very early universe. We point out that we are working with leading order solutions of asymptotic perturbation series and so, one cannot expect agreement beyond a certain time unless higher order term are added. Anyway, by a different choice of parameters, the range for the agreement can be extended. Fig.~\ref{fig-g} shows the result. The agreement is very good until time increases too much as expected.
we have chosen $g=100$ and $\lambda=1$ so to have $g\gg\lambda$ as expected from our analysis. Besides, we see that the gauge solution $f$ is negligible with respect to the scalar contribute by an order of magnitude with this parameters choice in agreement with our study.

To summarize, the scalar field interacts with a rapid varying gauge field that averages to a constant in the second of eqs.\ \eqref{eq:ff12}. This adds a shift to $\phi_0$ that is increasingly negligible at large times.

\medskip

\section{Evolution of cosmological quantities}\label{denergy}

\noindent Having obtained solutions in closed form for $\phi$ and $f$, we are in a position to map our analytical results with the numerical analysis performed in \cite{Rinaldi:2015iza,Alvarez:2019ues}. Our aim is to show that our solution can generically generate accelerating solutions. 

\begin{center}
\begin{figure}[H]
\includegraphics{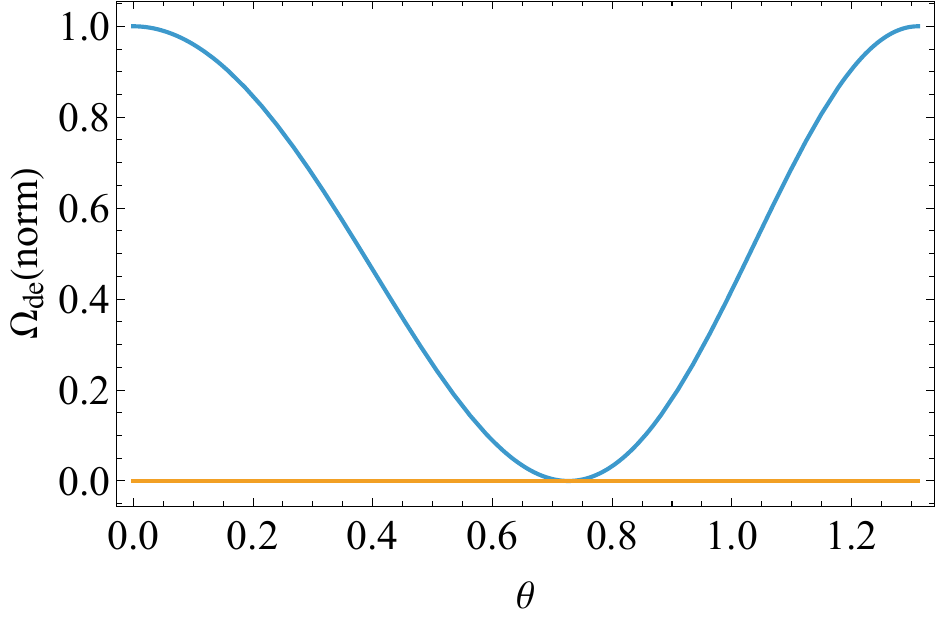}%
\caption{\it Dark energy density normalized to $\frac{\lambda \phi_0^4}{108M_p^2H^2}\approx 3\cdot 10^{53}$. From this plot it can be seen that infinite choices of $\theta$ exists granting agreement with data, being the Jacobi elliptic function periodic. \label{fig-3} 
}
\end{figure}
\end{center}

From the first Friedmann equation \eqref{Hsq} we define the critical density parameter for dark energy as
\bea \label{eqn1}
\Omega_{de}={1\over 3M_p^2H^2}\left[{\dot \phi^2\over 2}+{3\dot f^2\over 2a^2}+{2g^2f^4\over a^4}+{3g^2f^2\phi^2\over 8a^2}+V(\phi)\right]\,,
\eea 
that is we include all dynamical and potential contributions from the scalar and the gauge field. As stated above, we expect that the gauge contribution is effectively decoupled from gravity and does not contribute. Then, by assuming the time-scale of the Hubble constant to be the slowest one, we arrive, by means of eq.\ (\ref{eq:chi02}), to the following approximate density parameter for dark energy:
\bea\nonumber
&&\Omega_{de}=\frac{\lambda\phi_0^4}{108 H^2M^2}\left[3-2 \epsilon^2(t) \text{dn}^2\left(\theta +\frac{t \phi_0 \sqrt{\lambda }}{\sqrt{3}},-1\right)\right]^2+
\frac{4\lambda\phi_0^4}{108 H^2M^2}\epsilon^2(t)\times
\\
&&
\left[\text{cn}\left(\theta +\frac{t \phi_0 \sqrt{\lambda }}{\sqrt{3}},-1\right) \text{sn}\left(\theta +\frac{t \phi_0 \sqrt{\lambda }}{\sqrt{3}},-1\right)-3\frac{K_v}{\sqrt{2}}\frac{H}{\lambda\phi_0} \text{dn}\left(\theta +\frac{t \phi_0 \sqrt{\lambda }}{\sqrt{3}},-1\right)\right]^2, \nonumber \\
\eea
This expression is positive-definite by construction. To evaluate it, we need to extract the time variable through the scale factor given by \cite{Dutta:2008qn,Chiba:2009sj}
\be
a(t)=\sqrt[3]{\frac{1-\Omega_0}{\Omega_0}}\sinh^{\frac{2}{3}}\left(\frac{t}{t_\Lambda}\right),
\ee
where $\Omega_0$ is the present density of the scalar field,  $t_\Lambda=\sqrt{2/3\rho_{\phi_0}}$, and $\rho_{\phi_0}$ is the density of the scalar field. In our case, we assume that $\Omega_0$ tends to 0. Then, in the limit of large $t/t_\Lambda$ 
\be
\label{eq:Omega_de}
\Omega_{de}=\frac{\lambda \phi_0^4}{108M_p^2H^2}\left(2\ \text{dn}(\theta,-1)^2-3 \right)^2.
\ee
We plot the dark energy density parameter in Fig.~\ref{fig-3}.
This result grants the proper limit for $\Omega_{de}$. 

For completeness, we consider the other solution given in eq.(\ref{eq:phia}) to see what is the asymptotic limit in this case. This leads to
\be
\Omega_{de}=
\frac{1}{288M_p^2H^2}e^{-3Ht}\phi_0^2\left[18(-5+2\sqrt{6})H^2+((160-64\sqrt{6})+(49-20\sqrt{6})e^{-3Ht})\lambda\phi_0^2\right],
\ee
after averaging on the fast time scale determined by the period $T=2\pi/\sqrt{2\lambda\phi_0^2-9H^2/4}$. When $\lambda\phi_0^2\gg H^2$, this reduces to
\be
\Omega_{de}=e^{-3Ht}\frac{20-8\sqrt{6}}{3}\frac{H_L^2}{H^2},
\ee
where we have neglected the exponential $e^{-6Ht}$ with respect to the leading term.

\section{Comparison with DESI DR2 BAO data}

We use the BAO data from DESI DR2 to compare both at early universe (eq.(\ref{eq:chi02})) and at the late universe (eq.(\ref{eq:phia})). DESI data are those obtained through MCMC post-elaboration of the raw data. Specifically, we check for the CPL approximation in the plane $(w_0,w_a)$ and for the plot in the plane $(w_0,H_0)$.
\begin{center}
\begin{figure}[H]
\includegraphics[width=\textwidth]{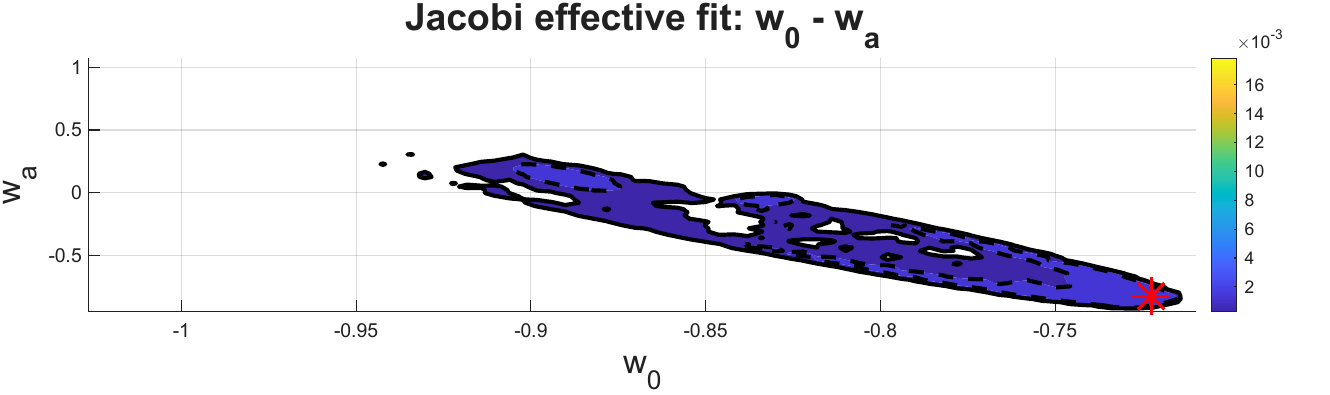}%
\caption{\it Early universe: The star is our model with the Jacobi elliptic functions. Curves represent 68\% and 95\% probabilities. The color bar on the right represents the local density of MCMC samples (i.e., how often the chain visits that region of parameter space). The agreement is rather good. \label{fig-4} 
}
\end{figure}
\end{center}

\begin{center}
\begin{figure}[H]
\includegraphics[width=\textwidth]{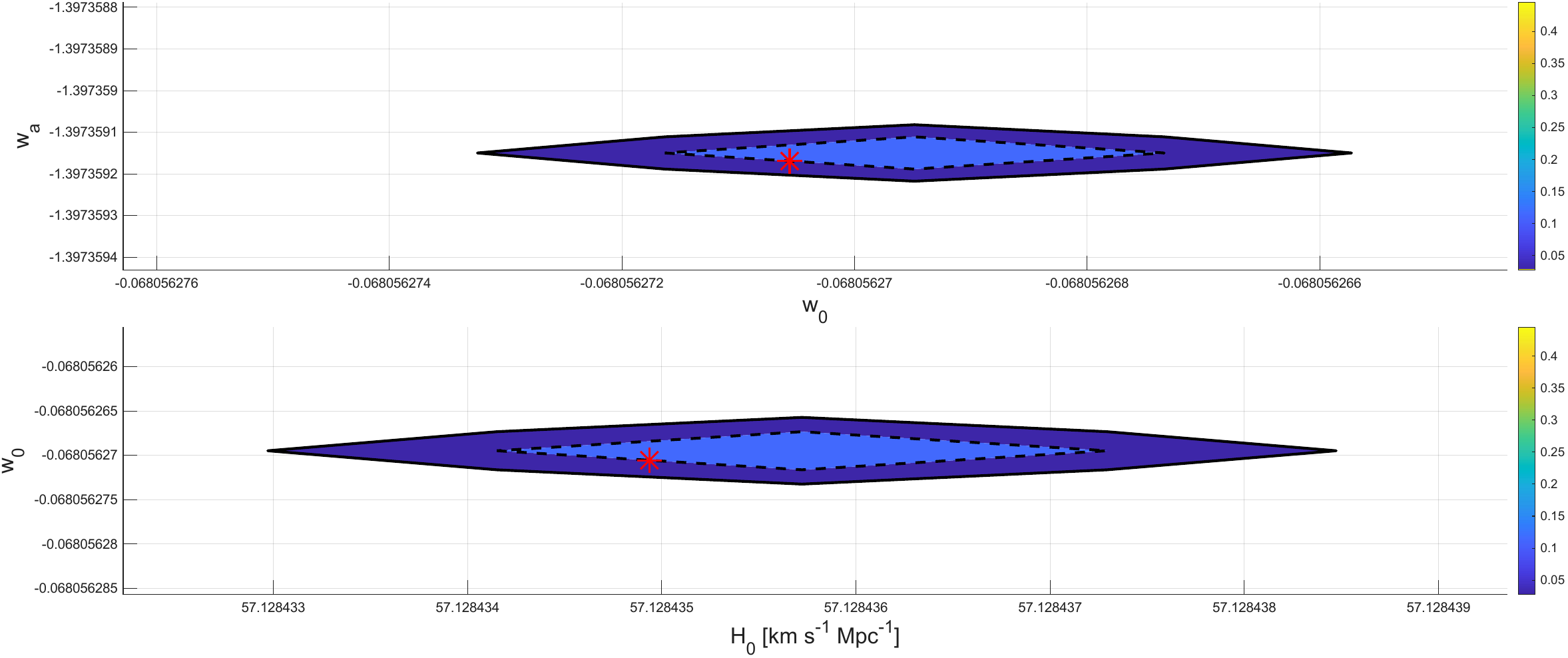}%
\caption{\it Late universe: The star is our model in eq.(\ref{eq:phia}). We performed 20000 runs. Curves represent 68\% and 95\% probabilities. The color bar on the right represents the local density of MCMC samples (i.e., how often the chain visits that region of parameter space). The agreement is very good. \label{fig-5} 
}
\end{figure}
\end{center}

In this case, the result is just approximate as eq.(\ref{eq:chi02}) can only apply to early universe. The situation is more interesting for eq.(\ref{eq:phia}). We performed an MCMC analysis to compare our model to DESI DR2 data and the result is given in Fig.~\ref{fig-5}.
It is important to emphasize that the agreement is reached after considering also $\phi_0$ as a parameter of the fit. This means that it cannot be taken the same as the v.e.v. of the electroweak breaking scale.

\section{Discussion and Conclusion}\label{conc}

\noindent To recall the main results, we derive closed analytical solutions for the equations of motion of the Yang-Mills--Higgs action for a dark sector (see eq.(\ref{eq:chi02}) and (\ref{eq:f0})) that permitted us to evaluate the equation of state for dark energy in eq.(\ref{eq:w}). The dark energy density was also evaluated in eq.(\ref{eq:Omega_de}). Therefore, we have shown that the classical Yang-Mills--Higgs action in the $SU(2)$ representation, and in a non-perturbative regime, can account for the current acceleration of the Universe, but also, more generally, to other accelerated expansion phases, such as inflation. We used a set of exact solutions for the Higgs and the Yang-Mills fields obeying a massive dispersion relation. Here, we summarize the salient features of our study:
\begin{itemize}
    \item Dark energy is a dynamical effect due to the asymptotic relaxation of the Higgs field on its vacuum expectation value. This yields a slowly varying effective cosmological constant, for instance $\theta$ in our case (see Fig. 3),  whose value can be set by tuning an integration constant, rather than physical constants. 
    \item The multi-scale analytic method employed here allows to overcome numerical ambiguities found in previous studies, based on dynamical systems methods, due to extreme sensitivity to the initial conditions, see e.g. \cite{Rinaldi:2015iza}.
    \item The possible interpretation of the current accelerated cosmic expansion as an effect of the interaction of  a Higgs-like field and gauge fields with gravity is natural in the framework of theories that consider "hidden" sectors of the Standard Model, see e.g. \cite{Patt:2006fw,Feldman:2007wj}.  This is true in view of the asymptotic solution we have found in eq.(\ref{eq:phia}) where the present value of the v.e.v. is reached without any slow-roll.
    \item A satisfactory comparison with DESI DR2 data, through MCMC testing, has been obtained for our model (see Fig.~\ref{fig-5}).
\end{itemize}
Summing up, in this study we have demonstrated that the system of EHYM equations involving Jacobi elliptical function leads generically to accelerated expansion. It leads to an equation of state parameter of order $-1$, depending on the values of a single parameter, that is the phase $\theta$. Further thorough and comprehensive analysis is required to verify if
this is truly viable, \textit{i.e.} if this leads to the currently observed distances and growth rates. However, this is  beyond the scope of the present manuscript and we leave it for future work. We envisage that such a non-perturbative framework will be very helpful in investigating cosmologies with strongly coupled theories in general which, so far, received less attention.

\section*{Acknowledgement}
The authors thanks Alexey Koshelev for fruitful discussions.

\section*{Funding and financial interests}

No funds, grants, or other support was received. The authors have no relevant financial or non-financial interests to disclose.

\section*{Data sharing}

Data sharing is not applicable to this article as no datasets were generated or analysed during the current study.

\section*{Appendix A: Review of Strongly Coupled Theory Treatment}

We summarize a brief review of the technique to compute Green's functin following Refs. \cite{Frasca:2015yva,Frasca:2015wva,Frasca:2013tma}.
For the following action
\be
S=\int d^4x\left(\frac{1}{2}(\partial\phi)^2-\frac{\lambda}{4}\phi^4+j\phi\right).
\ee
we have following equation of motion
\be
\label{eq:phi3}
-\Box\phi(x)=-\lambda\phi^3(x)+j(x).
\ee
In order to execute perturbation theory in the formal limit $\lambda\rightarrow\infty$. Rescaling $x\rightarrow\sqrt{\lambda}x$ and we put
\be
\phi(x)=\sum_{k=0}^\infty\lambda^{-k}\phi_k(x).
\ee
A direct substitution will yield the set of equations
\bea
-\Box\phi_0(x)&=&-\phi_0^3(x) \nonumber \\
-\Box\phi_1(x)&=&-3\phi_0^2(x)\phi_1(x)+j(x) \nonumber \\
-\Box\phi_2(x)&=&-3\phi_0^2(x)\phi_2(x)-3\phi_0(x)\phi_1^2(x) \nonumber \\
&\vdots&.
\eea
from which we see how one fixes the ordering for the source $j(x)$ \cite{Frasca:2009bc} and therefore, we do have a formal solution to the full set of perturbative equations.

For $\phi_n(x)$ we can obtain considering Eqn.(\ref{eq:phi3}) as a functional equation with $\phi=\phi[j]$ with the Taylor series expansion \cite{Frasca:2013tma}
\be
\phi[j]=\phi[0]+\int d^4y\left.\frac{\delta\phi}{\delta j(y)}\right|_{j=0}j(y)
+\frac{1}{2}\int d^4yd^4z\left.\frac{\delta^2\phi}{\delta j(y)\delta j(z)}\right|_{j=0}j(y)j(z)
+O(j^3).
\ee
we see we have got all the correlation functions for the classical theory. Now, for the 2-point function is
\be
G_c^{(2)}(x-y)=\left.\frac{\delta\phi}{\delta j(y)}\right|_{j=0},
\ee
and $\phi[0]=\phi_0(x)$. To avoid confusion, we will rename $\phi_0=\phi_c$ in the following. 

We call the 2-point function as the mass gap of the theory arises naturally, due to the presence of non-linear self-interaction
\be
\label{eq:phic}
\phi_c(x)=\mu\left(\frac{2}{\lambda}\right)^\frac{1}{4}{\rm sn}(p\cdot x+\theta,i)
\ee
where sn is a Jacobi elliptical function, $\mu$ and $\theta$ two arbitrary integration constants.  This holds provided the following relation is obeyed \cite{Frasca:2009bc}:
\be
\label{eq:DS}
p^2=\mu^2\sqrt{\frac{\lambda}{2}}.
\ee
We see we have got massive non-linear waves extending to all the space-time. 

The 2-point function in the momenta space becomes \cite{Frasca:2009bc}:
\begin{equation}
\label{eq:G1c}
    G_c^{(1)}(k)=\sum_{n=0}^\infty\frac{B_n}{k^2-m_n^2+i\epsilon}
\end{equation}
with
\begin{equation}
    B_n=(2n+1)^2\frac{\pi^2}{2K^2(i)}\frac{e^{-\left(n+\frac{1}{2}\right)\pi}}{1+e^{-(2n+1)\pi}}
\end{equation}
and
\begin{equation}
    m_n=(2n+1)\frac{\pi}{2K(i)}\left(\frac{\lambda}{2}\right)^\frac{1}{4}\mu.
\end{equation}
From this we can see that, in the limit $\mu\rightarrow 0$ the free theory is recovered at the leading order for the weak perturbation limit $\lambda\rightarrow 0$. 

For the Yang-Mills field, a similar argument applies. We consider a current expansion in the form \cite{Frasca:2023uaw}
\begin{equation}
A_\mu^{a}[j]=A_\mu^{a(0)}(x)+\sum_{k=1}^\infty\int
  C_{\mu_1\ldots \mu_k}^{a_1\ldots a_k(k)}(x,x_1,\ldots,x_k)
  \prod_{l=1}^kj^{a_l\mu_l}(x_l)d^4x_l,
\end{equation}
with the leading order equation
\begin{eqnarray}
\lefteqn{\partial^\mu(\partial_\mu A_\nu^{a(0)}-\partial_\nu A_\mu^{a(0)}
  +g\epsilon^{abc}A_\mu^{b(0)}A_\nu^{c(0)})}\nonumber\\&&\strut
  +g\epsilon^{abc}A^{b(0)\mu}(\partial_\mu A_\nu^{c(0)}
  -\partial_\nu A_\mu^{c(0)}+g\epsilon^{cde}A_\mu^{d(0)}A_\nu^{e(0)})=0,\qquad
\end{eqnarray}
and next-to-leading order is
\begin{eqnarray}
\lefteqn{\partial^2C_{\nu\rho}^{af(1)}(x,y)
  +g\epsilon^{abc}C_{\mu\rho}^{bf(1)}(x,y)\partial^\mu A_\nu^{c(0)}(x)
    +g\epsilon^{abc}A_\mu^{b(0)}(x)\partial^\mu C_{\nu\rho}^{cf(1)}(x,y)}
    \nonumber\\&&\strut
    +g\epsilon^{abc}C^{bf(1)\mu}_\rho(x,y)\partial_\mu A_\nu^{c(0)}(x)
    -g\epsilon^{abc}A^{b(0)\mu}(x)\partial_\nu C_{\mu\rho}^{cf(1)}(x,y)
    \nonumber\\&&\strut
    +g^2\epsilon^{abc}\epsilon^{cde}C_\rho^{bf(1)\mu}(x,y)A_\mu^{d(0)}(x)
    A_\nu^{e(0)}(x)+g^2\epsilon^{abc}\epsilon^{cde}A^{b(0)\mu}(x)
    C_{\mu\rho}^{df(1)}(x,y)A_\nu^{e(0)}(x)\nonumber\\&&\strut
    +g^2\epsilon^{abc}\epsilon^{cde}A^{b(0)\mu}(x)A_\mu^{d(0)}(x)
    C_{\nu\rho}^{ef(1)}(x,y)=\eta_{\nu\rho}\delta_{af}\delta^4(x-y).\qquad
\end{eqnarray}
To solve these equations, we use the mapping theorem proven in \cite{Frasca:2007uz,Frasca:2009yp} where the Yang-Mills potential can be written as
\be
A_\mu^a(x)=\eta_\mu^a\phi(x),
\ee
where $\eta_\mu^a$ are numerical coefficients with both color and Lorentz indexes and $\phi(x)$ satisfies the equation in the Lorentz gauge
\be
\Box\phi+Ng^2\phi^3=0,
\ee
for $SU(2)$. The mapping theorem turns back the Yang-Mills problem to the scalar field case discussed above. Then, also the equation for $C_{\mu\nu}^{ab(1)}(x,y)$ can be solved exactly yielding the same propagator as for the scalar field provided one accounts for the gauge projector and the colors contributing $\delta_{ab}$.

\section*{Appendix B: First order correction}

We have to solve the equation
\be
\phi_0^2\frac{\partial^2}{\partial t_\phi^2}\chi_1+(-\phi_0^2+3e^{-\tau}\chi_0^2)\chi_1+6H\phi_0\frac{\partial^2}{\partial t_\phi\partial\tau}\chi_0=0,
\ee
or, by using eq.(\ref{eq:chi01}),
\bea
&&\phi_0^2\frac{\partial^2}{\partial t_\phi^2}\chi_1+(-\phi_0^2+3e^{-\tau}\chi_0^2)\chi_1 \\
&&+\sqrt{2}H\phi_0^2e^{\tau/2}
\operatorname{sn}\left(\frac{t_\phi}{\sqrt{3}}+\theta,-1\right)\operatorname{cn}\left(\frac{t_\phi}{\sqrt{3}}+\theta,-1\right)=0.
\eea

To solve this equation, we need the Green function that solves the equation
\be
\frac{\partial^2}{\partial t^2}G(t,t')+\lambda(-\phi_0^2+3e^{-\tau}\chi_0^2)G(t,t')=\delta(t-t'),
\ee
that is solved by
\be
G(t,t')=\frac{\sqrt{3}}{\sqrt{\lambda}\phi_0}\theta(t-t')\operatorname{sn}\left(\sqrt{\frac{\lambda}{3}}\phi_0(t-t'),-1\right)\operatorname{cn}\left(\sqrt{\frac{\lambda}{3}}\phi_0(t-t'),-1\right).
\ee
Thus, we get
\bea
&&\chi_1=-\sqrt{6}H\phi_0e^{\tau/2}\int_0^t dt'
\operatorname{sn}\left(\sqrt{\frac{\lambda}{3}}\phi_0(t-t'),-1\right)\operatorname{cn}\left(\sqrt{\frac{\lambda}{3}}\phi_0(t-t'),-1\right)\times \nonumber \\
&&\operatorname{sn}\left(\sqrt{\frac{\lambda}{3}}\phi_0t'+\theta,-1\right)\operatorname{cn}\left(\sqrt{\frac{\lambda}{3}}\phi_0t'+\theta,-1\right).
\eea
We now use the Fourier series for the Jacobi functions. Firstly, one observe that $(\operatorname{dn}(x,-1))'=\operatorname{sn}(x,-1)\operatorname{cn}(x,-1)$ and we take
\be
\label{eq:dn}
\operatorname{dn}(x,-1)=\frac{\pi}{2K(-1)}+\frac{2\pi}{K(-1)}\sum_{n=1}^\infty(-1)^n\frac{e^{-n\pi}}{1+e^{-2n\pi}}\cos\left(\frac{n\pi}{K(-1)}x\right).
\ee
Thus,
\be
(\operatorname{dn}(x,-1))'=-\frac{2\pi^2}{K^2(-1)}\sum_{n=1}^\infty(-1)^nn\frac{e^{-n\pi}}{1+e^{-2n\pi}}\sin\left(\frac{n\pi}{K(-1)}x\right)=\operatorname{sn}(x,-1)\operatorname{cn}(x,-1).
\ee
We have to evaluate
\bea
&&\chi_1=-\sqrt{6}H\phi_0e^{\tau/2}\frac{4\pi^4}{K^4(-1)}\sum_{m=1}^\infty\sum_{n=1}^\infty(-1)^{m+n}mn\frac{e^{-(m+n)\pi}}{(1+e^{-2m\pi})(1+e^{-2n\pi})} \nonumber \\
&&\int_0^t dt'\sin\left(\frac{m\pi}{K(-1)}\sqrt{\frac{\lambda}{3}}\phi_0(t-t')\right)\sin\left(\frac{n\pi}{K(-1)}\left(\sqrt{\frac{\lambda}{3}}\phi_0t'+\theta\right)\right).
\eea
For $m=n$, we get a secular term given by
\be
6\phi_0e^{\tau/2}\frac{2\pi^4}{K^4(-1)}\left(-\frac{H}{\sqrt{6}}t\right)\sum_{n=1}^\infty n^2\frac{e^{-2n\pi}}{(1+e^{-2n\pi})^2}\cos\left(\frac{n\pi}{K(-1)}\left(\sqrt{\frac{\lambda}{3}}\phi_0t+\theta\right)\right),
\ee
that is unbounded in the asymptotic limit. 
To remove the secularity we need the envelope method \cite{Kunihiro:1995zt}. We write our solution in the form
\bea
&&\chi=e^{\frac{\tau}{2}}\sqrt{\frac{2}{3}}\phi_0\,\text{dn}\left(\sqrt{\frac{\lambda}{3}}\phi_0t+\theta,-1\right)+ \\
&&6\lambda^{-\frac{1}{2}}\phi_0e^{\tau/2}\frac{2\pi^4}{K^4(-1)}\left(-\frac{H}{\sqrt{6}}(t-t_0)\right)\sum_{n=1}^\infty n^2\frac{e^{-2n\pi}}{(1+e^{-2n\pi})^2}\cos\left(\frac{n\pi}{K(-1)}\left(\sqrt{\frac{\lambda}{3}}\phi_0t+\theta\right)\right) 
+r.t.,  \nonumber
\eea
where $r.t.$ are regular (no secular) terms and we have introduced an initial time $t_0$ while we assume that $\phi_0$ is also a function of $t_0$ and assume $\theta=-\sqrt{\frac{\lambda}{3}}\phi_0t_0$. Then, we use eq.(\ref{eq:dn}) to get
\bea
&&\chi=e^{\frac{\tau}{2}}\sqrt{\frac{2}{3}}\phi_0\left[\frac{\pi}{2K(-1)}+\frac{2\pi}{K(-1)}\sum_{n=1}^\infty(-1)^n\frac{e^{-n\pi}}{1+e^{-2n\pi}}\cos\left(\frac{n\pi}{K(-1)}\left(\sqrt{\frac{\lambda}{3}}\phi_0(t-t_0)\right)\right)\right]+\\
&&6\lambda^{-\frac{1}{2}}\phi_0e^{\tau/2}\frac{2\pi^4}{K^4(-1)}\left(-\frac{H}{\sqrt{6}}(t-t_0)\right)\sum_{n=1}^\infty n^2\frac{e^{-2n\pi}}{(1+e^{-2n\pi})^2}\cos\left(\frac{n\pi}{K(-1)}\left(\sqrt{\frac{\lambda}{3}}\phi_0(t-t_0)\right)\right)
+r.t. \nonumber
\eea
The condition for the envelope is \cite{Kunihiro:1995zt} 
\be
\left.\frac{d\chi}{dt_0}\right|_{t_0=t}=0,
\ee
that yields
\be
\sqrt{\frac{2}{3}}\frac{d\phi_0}{dt}+K_v\lambda^{-\frac{1}{2}}\phi_0H=0,
\ee
where use has been made of the identity
\be
\frac{\pi}{2K(-1)}+\frac{2\pi}{K(-1)}\sum_{n=1}^\infty(-1)^n\frac{e^{-n\pi}}{1+e^{-2n\pi}}=1,
\ee
and
\be
K_v=\frac{6\pi^4}{K^4(-1)}\sum_{n=1}^\infty n^2\frac{e^{-2n\pi}}{(1+e^{-2n\pi})^2}=0.37084\ldots.
\ee
Thus,
\be
\phi_0(\tau)=e^{-\sqrt{\frac{1}{6\lambda}}K_v\tau}\phi_0(0).
\ee
This correction must apply only to the amplitude of $\chi$ and not to the phase to grant that no other secular terms appears without control putting back the solution into the equation.

\end{document}